\documentclass[twocolumn,prl,unsortedaddress,superscriptaddress]{revtex4}
\usepackage{amsmath,color}
\usepackage[dvips]{graphicx}
\usepackage[scaled]{uarial}
\usepackage[T1]{fontenc}

\allowdisplaybreaks

\usepackage{lipsum}
\begin{document}
\title{Origin of Negative Density and Modulus in Acoustic Metamaterials}

\vspace{\baselineskip}

\author{Sam H. Lee\footnote{e-mail:samlee@yonsei.ac.kr}}
\affiliation{Institute of Physics and Applied Physics, Yonsei
University, Seoul 120-749, Korea}
\author{Oliver B. Wright}
\affiliation{Division of Applied Physics, Faculty of Engineering, Hokkaido University,
Sapporo 060-8628, Japan}


\vspace{\baselineskip}

\begin{abstract}
This paper provides a review and fundamental physical interpretation for the effective densities and moduli of acoustic metamaterials. We introduce the terminology of hidden force and hidden source of volume: the
 effective density or modulus is negative when the hidden force or source of volume is larger than, and operates in antiphase to, respectively, the force or volume change that would be obtained in their absence. We demonstrate this ansatz for some established acoustic metamaterials with elements based on membranes, Helmholtz resonators, springs and masses. The hidden force for membrane-based acoustic metamaterials, for instance, is the force from the membrane tension. The hidden source for a Helmholtz-resonator-based metamaterial is the extra air volume injected from the resonator cavity. We also explain the analogous concepts for pure mass-and-spring systems, in which case hidden forces can arise from masses and springs fixed inside other masses, whereas hidden sources\textemdash more aptly termed hidden expanders of displacement in this case\textemdash can arise from light rigid trusses coupled to extra degrees of freedom for mechanical motion such as the case of coupling to masses that move at right angles to the wave-propagation direction. This overall picture provides a powerful tool for conceptual understanding and design of new acoustic metamaterials, and avoids common pitfalls involved in determining the effective parameters of such materials.
\end{abstract}

\maketitle

Acoustic metamaterials are man-made structures designed
to manipulate the propagation of sound in ways not available in naturally occurring materials.
The understanding of negative constitutive parameters in such materials,\cite{liu2000locally,li2004double,fang2006ultrasonic,ding2007metamaterial,yang2008membrane,cheng2008one,ao2008far,lee2009density,lee2009acoustic,ding2010two,lee2010composite,christensen2010all,liu2011elastic,wu2011elastic,fok2011negative,lai2011hybrid,liang2012extreme,garcia2012quasi,seo2012acoustic,lee2012highly,park2013giant,chen2013double,yang2013coupled,zhai2013double,liang2013space,xie2013measurement,garcia2014negative,norris2014metal,xie2014compressive,koju2014extraordinary,crow2015experimental,brunet2015soft,park2015acoustic,popa2015water} one of their most exotic features, has to date relied more on engineering-based concepts than on universal physical principles. This can lead to confusion in assigning effective parameters to a given system.
Here we present a fundamental physical picture to account for the effective densities and moduli of acoustic metamaterials that allows their intuitive yet precise understanding, and at the same time allows their unambiguous determination.
This picture is based on hidden forces and hidden sources of volume. The former, resulting in an effective mass or density, involve local, time-dependent non-apparent forces that provide a net force on the unit cell of the acoustic metamaterial. The latter, resulting in an effective elastic modulus, involve time-dependent hidden sources of volume or displacement that only produce pairs of forces acting equally and oppositely on either side of the unit cell.

 We first illustrate our approach with examples of systems exhibiting an effective mass based on membranes in tubes as well as on mass-and-spring models (the latter providing elements to model three-dimensional (3D) solid acoustic metamaterials). An example of a metamaterial based on  membranes combined with masses and springs is also elucidated. We then provide examples of systems exhibiting an effective modulus based on Helmholtz-resonators in tubes as well as based on mass-and-spring models combined with light rigid trusses coupled to extra degrees of freedom for mechanical motion. Systems exhibiting both effective density and modulus, including the possibility of double-negative parameter behavior, are also discussed.

\section{Effective densities: hidden forces}

In this section we introduce the concept of ``hidden force'' in order to understand effective mass in systems that contain non-apparent mechanical elements.\cite{milton2007modifications} This approach is first explained by means of simple examples from mechanics. We then illustrate the concept of effective density for several different acoustic metamaterials involving unidirectional propagation: a membrane-based metamaterial, a solid-matrix metamaterial approximated by spring-coupled masses, and also a new membrane-based metamaterial that also includes masses and springs.

\subsection{Simple mechanical systems}

Consider a wheel of mass $M$, radius $R$ and  moment of
inertia $I$ rolling in the $x$ direction without slipping on a flat, horizontal surface, as shown in Fig. 1(a). A horizontal force $F$ is applied to the axis of the
wheel at its center. The application of Newton's laws allows one to derive the acceleration: $\ddot{x}  = F/[M(1+I/MR^2])$. The effective
mass of wheel can be defined as
\begin{equation} \label{one}
M_{eff}=F/\ddot{x} ,
\end{equation}
giving, in this case, $M_{eff} = M(1+I/MR^2)$. Let us define the hidden force $F_h$ to be the backwardly-directed frictional
force on the wheel rim. The acceleration $\ddot{x}$ is smaller than that expected for a non-rotating mass of the same magnitude, i.e. $\ddot{x}<F/M$. By
introducing the effective mass as so defined, it is thus possible to obtain the correct acceleration from one simple equation.

\begin{figure}
\begin{center}
\includegraphics*[width=1.0\columnwidth]{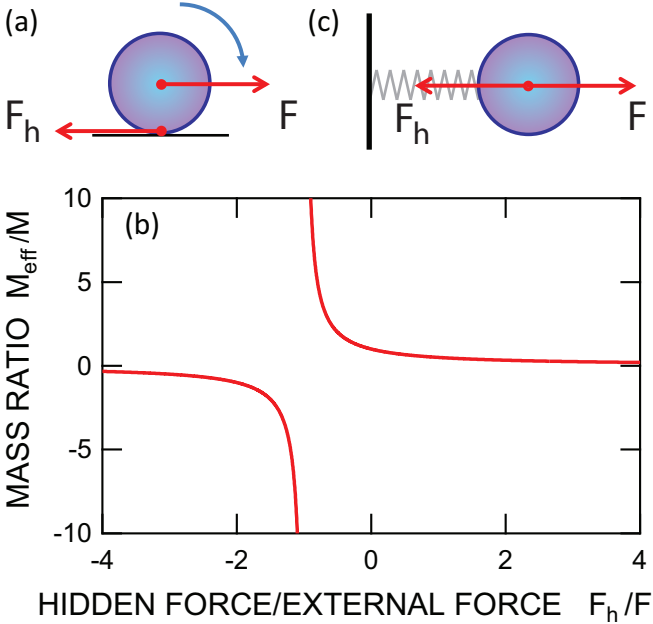}
\end{center}
\caption{(a) Showing external and hidden forces, $F$ and $F_h$, for the case of a rolling wheel. (b) shows a normalized plot of the effective mass as a function of the hidden force. (c) shows $F$ and $F_h$ for a prototype of a system exhibiting negative effective mass: a simple harmonic oscillator.} \label{fig:set1}
\end{figure}

In general, if an external force $F$ is applied to a mass $M$, then, owing to the specific mechanism involved, a hidden force $F_h$ may also act on the mass (assumed here to be collinear with $F$). So the acceleration becomes $\ddot{x}=(F+F_h)/M$, and the effective mass is given by
\begin{equation} \label{two}
M_{eff}= \frac{M}{1+F_h/F}.
\end{equation}
A plot of $M_{eff}/M$ vs $F_h/F$ is shown in Fig. 1(b). Notably, as $F_h$ approaches $-F$, $M_{eff}$ becomes
infinitely large. Also, $M_{eff}$ becomes negative when $F_h<-F$.

A prototype of a system exhibiting negative $M_{eff}$ is a simple harmonic oscillator consisting of a mass $M$ attached to a rigid wall by
a spring, as shown in Fig. 1(c): the hidden-force picture starts by ignoring the presence of the spring and regarding the system as a free mass subject to a hidden
force $F_h=-kx$, where $k$ is the spring constant and $x$ the displacement. In the case when the external driving force is sinusoidal at angular frequency $\omega$, 
$F=F_0\exp (-i\omega t)$, the acceleration can be calculated from the equation of
motion, $M\ddot{x}=F_0 \exp(-i\omega t)+F_h$.
Using the harmonic expression $x=x_0\exp (-i\omega t)$, we then obtain the hidden force 
$F_h=-kx=-F \omega^2_0/(\omega^2_0-\omega^2)$, where $\omega_0=\sqrt{k/M}$ is the resonance
frequency. Substituting into Eq. (\ref{two}),
\begin{equation} \label{three}
M_{eff}= M\left( 1-\frac{\omega^2_0}{\omega^2}\right).
\end{equation}
The displacement, obtained by integrating $\ddot{x}=F/M_{eff}$, can be clearly seen to oscillate with large amplitude near resonance
(at $\omega_0$) because $M_{eff}$ becomes very small. This ansatz describes all physical
quantities correctly and quantitatively. Another example is the abrupt shift of the phase of the displacement by $\pi$ with respect to the driving force as
the frequency passes through the resonance. This can immediately be understood from Eq.
(\ref{three}), since the sign of $M_{eff}$ changes at $\omega =\omega_0$. Negative $M_{eff}$, a consequence of $F_h<-F$, implies in the case of sinusoidal excitation that the magnitude of the hidden force is not only in antiphase with but also has a magnitude that is larger than that of the applied force. As the limit $\omega=0$ is approached, the effective mass tends to $-\infty$ because the required force $F$ for a given oscillation amplitude becomes tiny in comparison with the oppositely directed spring force $F_h$. As the limit $\omega=\infty$ is approached, the effective mass tends to $M$ because the (inertial) force $F$ required for a given oscillation amplitude becomes very large in comparison with the spring force $F_h$.
\subsection{Membrane-based acoustic metamaterial}

\begin{figure*}
\begin{center}
\includegraphics*[width=2.0\columnwidth]{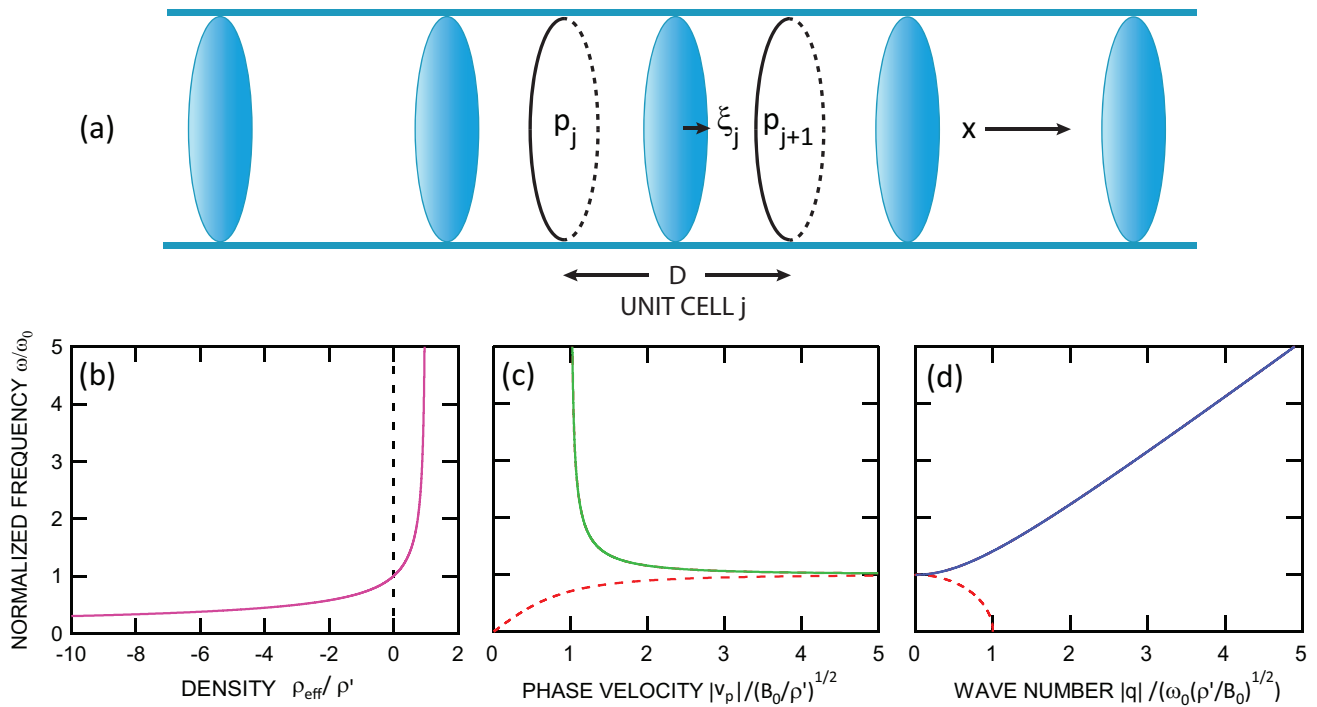}
\end{center}
\caption{(a) Schematic diagram of a 1D
membrane-based acoustic metamaterial, made up of periodically-spaced membranes in a tube containing air. Normalized plots (b), (c), (d) show the frequency as a function of the density, phase velocity and wave number, respectively. In (c) and (d) the solid and dashed lines refer to the cases of real and imaginary values for the phase velocity and wave number, respectively.} \label{fig:set2}
\end{figure*}

Acoustic metamaterials, consisting of arrays of resonators, naturally fit into this hidden-force picture. Take the example of a 1D 
membrane-based acoustic metamaterial,\cite{yang2008membrane,lee2009acoustic,lee2009density,lee2010composite,naify2010transmission,mei2012dark,yang2013coupled,park2015acoustic} which supports wave propagation down its length, as shown schematically in Fig. 2(a). It consists of a cylindrical air-filled tube containing taut membranes at regular intervals. Consider a particular unit cell. Its
center of mass $M$ is subject to two kinds of forces: the applied force $F=S\Delta p$ from the two adjacent cells, where $S$ and $\Delta p$ are the cross-sectional area of
the tube and the pressure difference across the unit cell. The hidden force is $F_h=-k_m\xi$, where $k_m$ and $\xi$ are the membrane spring constant\cite{lee2009density} and the displacement of the unit-cell center of
mass.\footnote{We are in fact making the approximation that the air in the unit cell on both sides of the membrane moves together with the membrane, the so-called lumped-element approach.\cite{blackstock2000fundamentals} Practically, this approach cannot be exact because strictly one should solve for the acoustic particle velocity at every point inside the tube. However, the lumped-element approach, based on unit cells much smaller than the acoustic wavelength $\lambda$, represents an excellent approximation.} The equations governing oscillatory motion are the same as those governing the system of a mass connected by a spring to a rigid wall that we just treated: $M\ddot{\xi}=S\Delta p+F_h$, where $M=\rho_0 SD+M_m$ is the mass of a unit cell. Here, $\rho_0$ is the density of air, $D$ is the unit-cell length and $M_m$ is the mass of the membrane. For sinusoidal excitations, $F=S\Delta p=M\ddot{\xi}+k_m\xi=\ddot{\xi}(M-k_m/\omega^2)$, so $\ddot{\xi}=F/[M(1-\omega_0^2/\omega^2)]$, where $\omega_0=\sqrt{k_m/M}$.

According to Eq. (\ref{one}), the unit-cell effective mass is $M_{eff}=M(1-\omega^2_0/\omega^2)$, which has exactly the same form as $M_{eff}$ for a simple harmonic oscillator. This treatment, based on lumped-elements, obviously ignores vibrational resonances of the membranes higher than the fundamental mode.\cite{blackstock2000fundamentals}

For each unit cell indexed by $j$, the effective mass can be defined as
\begin{equation} \label{five1}
M_{eff}=\frac{F_j-F_{j+1}}{\ddot{\xi}_j}=\frac{S(p_j-p_{j+1})}{\ddot{\xi}_j},
\end{equation}
where $F_j$ is the force acting on the left-hand side of the unit cell $j$, whereas $-F_{j+1}$ is that acting on the right-hand side. (Alternatively, if we define $f_j$ as the force on the left-hand side and $+f_{j+1}$ as the force on the right, we obtain $M_{eff}=(f_j+f_{j+1})/\ddot{\xi}_j$, which demonstrates more clearly that $F=F_j-F_{j+1}=f_j+f_{j+1}$ is the net force on unit cell $j$. However, in this paper we adopt the $F_j$ notation because it is more convenient for the analysis of mass-and-spring models.) The effective density can be defined by $\rho_{eff}=M_{eff}/V$, where $V=SD$ is the volume of the unit cell, which is a useful concept for the case $D\ll\lambda$, where $\lambda$ is the acoustic wavelength. It is this limit that applies to metamaterials, as opposed to the case for phononic crystals for which $D\sim\lambda$. Therefore
\begin{equation} \label{seven}
\rho_{eff}=\rho'\left(1-\frac{\omega^2_0}{\omega^2}\right),
\end{equation}
where the average density $\rho'$ is given by $\rho'=M/V$.

In order to derive the system's effective bulk modulus $B_{eff}$, consider the volume $V$ between two adjacent membranes comprising parts of both the $j$th and ($j-1$)th unit cells. The non-equilibrium component of pressure in this volume, $p_j$,\footnote{Strictly speaking, the pressure in this volume is the average of the pressures $p_j$ and $p_{j+1}$ at the two sides of the unit cell $j$, but we approximate here to $p_j$ as the small difference does not change the final result.} is related in general to the deviation $\Delta V_j$ (which we define to refer to this same volume) from the equilibrium volume $V$ as follows:
\begin{equation} \label{eight}
p_j=-B_{eff} \frac{\Delta V_j}{V}=-B_{eff} \frac{S(\xi_{j}-\xi_{j-1})}{SD},
\end{equation}
where $B_{eff}$ is the effective modulus of the system.\footnote{As mentioned, $S(\xi_{j}-\xi_{j-1})$ is the deviation from the equilibrium volume averaged over two adjacent unit cells, but the small difference between this quantity and the more exact expression $S(\xi_{j+1}-\xi_{j-1})/2$ can be neglected.} In the present geometry, only the (adiabatic) bulk modulus $B_0$ of the air in the tube affects the pressure-volume relation,
\begin{equation} \label{eight2} 
p_j=-B_0\frac{\Delta V_j}{V}, 
\end{equation}
and so $B_{eff}=B_0$. Equation (\ref{eight}) represents continuity combined with the equation of state, and can be rewritten in the form
\begin{equation} \label{three1}
B_{eff}^{-1}=-\frac{1}{p_j}\frac{\Delta V_j}{V}=-\frac{1}{p_j}\frac{\xi_{j}-\xi_{j-1}}{D}.
\end{equation}
In spite of the introduction of membranes in the tube with spring-like properties, they evidently do not change the effective modulus $B_0$ of this acoustic metamaterial. 

One can also access the acoustic dispersion relation as follows. In the metamaterial limit, i.e. $D\ll\lambda$, the time derivative of Eq. (\ref{eight}) leads, for an arbitrary point in the tube, to
\begin{equation} \label{ten}
\dot{p}=-B_0\frac{\partial u}{\partial x}.
\end{equation}
where $u=\dot{\xi}$ is the acoustic particle velocity and $x$ is the distance along the tube. From the definition $\rho_{eff}=M_{eff}/(AD)$ and Eq. (\ref{five1}), we also have
\begin{equation} \label{six}
-\frac{\partial p}{\partial x}=\rho_{eff}\dot{u}.
\end{equation}
The wave equation is easily obtained by combining Eqs. (\ref{ten}) and (\ref{six}):
\begin{equation} \label{eleven}
\rho_{eff}\ddot{p}=B_0\frac{\partial^2 p}{\partial x^2}.
\end{equation}
Substituting $p=p_0\exp[i(qx-\omega t)]$, where $q$ is the wave number, leads to the dispersion relation $q^2=\omega^2\rho_{eff}/B_0$. For the example in question,
\begin{equation} \label{twelve}
q=\omega \sqrt{\frac{\rho_{eff}}{B_0}}=\sqrt{\frac{\rho'}{B_0}}\left(1-\frac{\omega^2_0}{\omega^2}\right)^{1/2}.
\end{equation}
A plot of the frequency dependence of the effective density, and also plots for $|v_p|$,  where $v_p$ is the phase velocity ($v_p=\omega/q$) and $|q|$, where $q$ is the wave number, are shown on normalized scales in Fig. 2(b)-(d). The phase velocity becomes infinite when the effective density vanishes at $\omega=\omega_0$. This situation is useful for applications in extraordinary acoustic transmission.\cite{park2013giant,fleury2013extraordinary} At lower frequencies, where $\rho_{eff}$ is negative, the waves are damped [Im$(q)>0$, dashed lines in (c) and (d)], whereas at higher frequencies, where the $\rho_{eff}$ is positive, they are undamped [Im$(q)=0$, solid lines in (c) and (d)]. The frequency region of damping, sometimes referred to as a metamaterial band gap, is useful in practice for applications to the absorption of noise.

One of the most curious features of these results is how membranes with spring-like properties only contribute to the effective density, and do not influence effective modulus. We shall see later that this can also be explained by the fact that hidden sources of volume rather than hidden forces influence the effective modulus.

\begin{figure*}
\begin{center}
\includegraphics*[width=2.0\columnwidth]{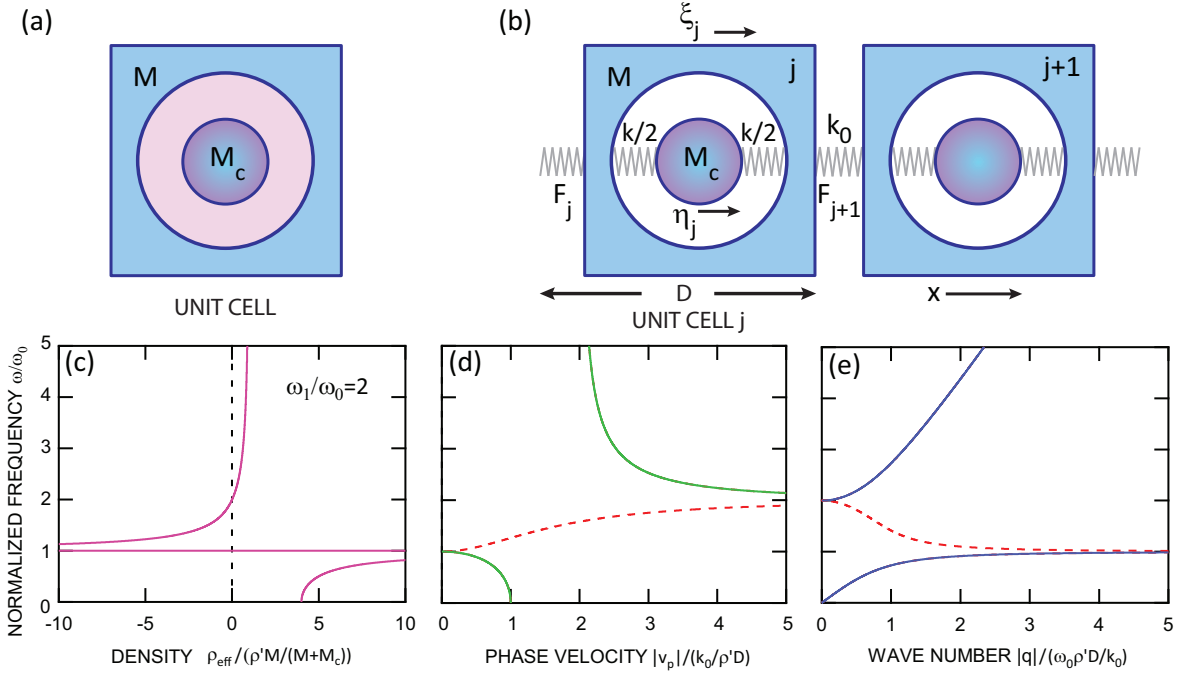}
\end{center}
\caption{(a) Unit cell of a solid-matrix acoustic metamaterial, modelled by a spherical core of mass $M_c$ surrounded by a separate, concentric rigid shell of mass $M$. (b) mass-and-spring analogy of a 1D chain of unit cells. $F_j$ is the compressional force in the spring of the left-hand side of unit cell $j$. Normalized plots (c), (d), (e) show the frequency as a function of the density, phase velocity and wave number, respectively. In (d) and (e) the solid and dashed lines refer to the cases of real and imaginary values, respectively, for the phase velocity and wave number. Normalized parameter $\omega_1/\omega_0=2$ is chosen for these plots.} \label{fig:set3}
\end{figure*}

\subsection{Mass-and-spring analogy for a solid-matrix acoustic metamaterial}

Now consider a metamaterial composed of a cubic array of mechanical resonators embedded in a compliant solid matrix, as illustrated by the unit cell in Fig. 3(a). Such a system consisting of rubber-coated lead balls was reported to exhibit strong sonic
transmission loss due to negative density.\cite{liu2000locally}
Consider a 1D mass-and-spring analog of such a matrix in the form of a chain,\cite{huang2012anomalous} as shown in Fig. 3(b). The unit cell can be modelled by a core of mass $M_c$ surrounded by a hollow mass $M$, with an internal connection between the masses made up of collinear springs of constant $k/2$. Springs of constant $k_0$ connect the masses $M$ externally. The external springs transmit applied forces $F$ to a particular unit cell, where
$F=F_0\exp(-i\omega t)$, whereas
the hidden force $F_h=-k(\xi-\eta)$ acts on mass $M$ through the internal springs, where $\xi$ and $\eta$ are the displacements of the masses $M$ and
$M_c$, respectively. Equations of motion for these parameters are $M\ddot{\xi}+k(\xi-\eta)=
F_0\exp(-i\omega t)$ and $M_c\ddot{\eta}+k(\eta-\xi)=0$ respectively. This clearly shows the
origin of the hidden force as the vibration of the mass $M_c$. Solving the coupled equations for sinusoidal motion, we
obtain the effective mass of a unit cell in the form
\begin{equation} \label{thirt}
M_{eff}=F/\ddot{\xi}=M+\frac{M_c}{1-\omega^2/\omega_0^2}=M\left(1+\frac{\omega_1^2-\omega_0^2}{\omega_0^2-\omega^2}\right),
\end{equation}
where $\omega_1=\omega_0\sqrt{1+M_c/M}$ and $\omega_0=\sqrt{k/M_c}$ (distinct from $\omega_0$ in the above membrane problem). A plot of the normalized value of $M_{eff}$ (i.e. normalized effective density) vs frequency is shown in Fig. 3(c), for the case $\omega_1/\omega_0=2$. One can see that as the frequency increases through $\omega=\omega_0$, there is a transition from infinitely positive to infinitely negative $M_{eff}$. Zero $M_{eff}$ occurs at $\omega=\omega_1$.

The effective modulus can be derived in a similar way to that for the membrane-based metamaterial. By analogy with Eq. (\ref{eight}),
\begin{equation} \label{eight1}
F_j=-k_{eff}(\xi_j-\xi_{j-1})=-D^2E_{eff} \frac{\xi_{j}-\xi_{j-1}}{D},
\end{equation}
where $k_{eff}$ is an effective spring constant and $E_{eff}$ the effective Young's modulus of the system. Alternatively,
\begin{equation} \label{fourt1}
E_{eff}^{-1}=-\frac{D}{F_j}(\xi_{j}-\xi_{j-1}).
\end{equation}
For the present 1D mechanical model, the compressive force $F_j$ is provided by the spring of constant $k_0$, i.e. $F_j=-k_0(\xi_{j}-\xi_{j-1})$, so $k_{eff}=k_0$ and $E_{eff}=k_0/D$. Clearly, the effective modulus of this structure not affected by the internal structure of the mass. As in the case of the membrane-based metamaterial, the effective modulus is positive and frequency independent.

The acoustic dispersion relation can be derived by application of Newton's second law for sinusoidal variations:
\begin{equation} \label{fourt}
M_{eff}\ddot{\xi_j}=k_0(\xi_{j-1}-\xi_{j})-k_0(\xi_j-\xi_{j+1})\approx k_0D^2\frac{\partial^2 \xi_j}{\partial x^2},
\end{equation}
where we have assumed, as before, that $D\ll\lambda$.
Making use of $\rho_{eff}=M_{eff}/D^3$ and $E_{eff}=k_0/D$, we may write
\begin{equation} \label{fift}
\rho_{eff}\ddot{\xi}=E_{eff}\frac{\partial^2 \xi}{\partial x^2},
\end{equation}
where
\begin{equation} \label{sixt}
\rho_{eff}=\rho'\frac{M}{M+M_c}\left(1+\frac{\omega_1^2-\omega_0^2}{\omega_0^2-\omega^2}\right).
\end{equation}
The constant $\rho'=(M+M_c)/D^3$ is the average
density of the matrix. 
Substituting $\xi=\xi_0\exp[i(qx-\omega t)]$ leads to the dispersion relation $q=\omega\sqrt{\rho_{eff}/E_{eff}}$, or
\begin{equation} \label{sixtz}
q=\omega\frac{\omega_0}{\omega_1}\sqrt{{\frac{\rho'D}{k_0}}}\left(1+\frac{\omega_1^2-\omega_0^2}{\omega_0^2-\omega^2}\right)^{1/2}.
\end{equation}
The frequency dependence of the wave number together with those of the phase velocity and effective density are plotted on normalized scales in Figs. 3(d) and (e) for the case $\omega_1/\omega_0=2$.
In the frequency range $\omega_0 < \omega < \omega_1$, $\rho_{eff}$ is
negative and the waves are damped. Because this system has more degrees of freedom than the membrane system previously discussed, it has a more complicated dispersion relation.

\begin{figure*}
\begin{center}
\includegraphics*[width=2.0\columnwidth]{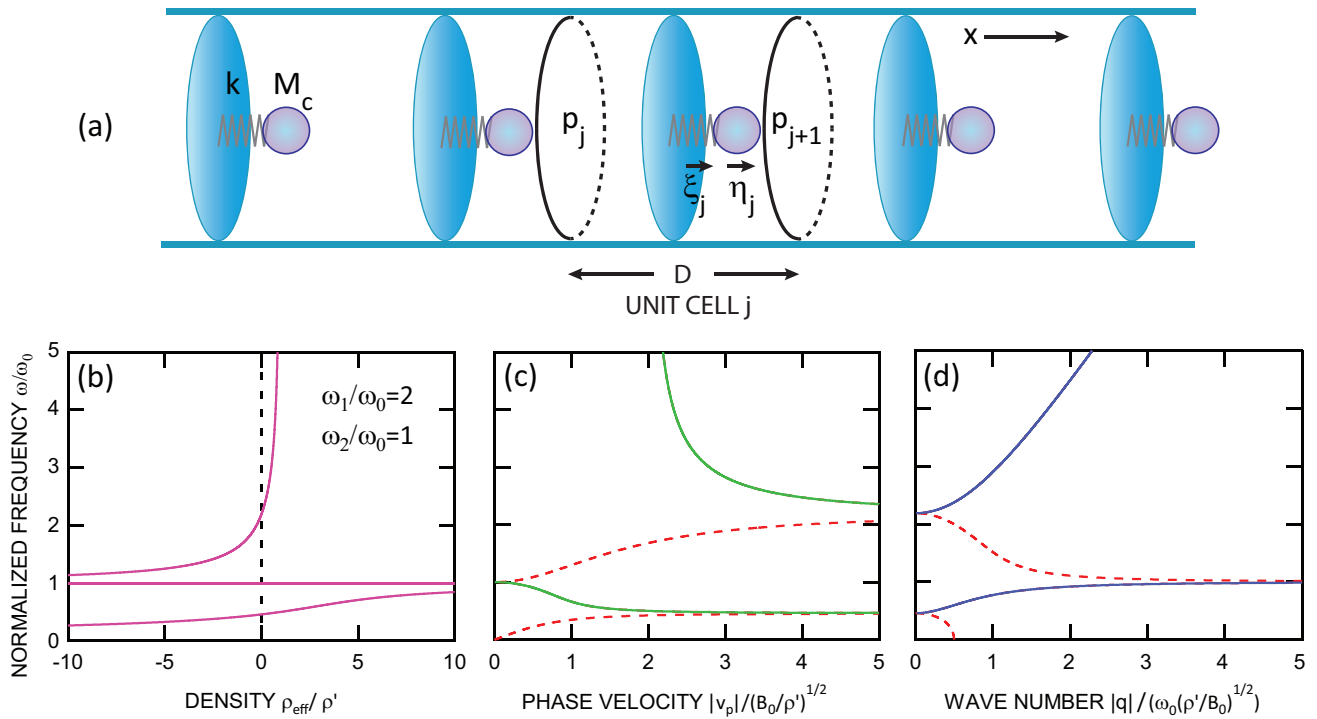}
\end{center}
\caption{(a) Schematic diagram of a 1D
acoustic metamaterial based on membranes, springs and masses, made up of periodically-spaced unit cells in a tube containing air. Normalized plots (b), (c), (d) show the frequency as a function of the density, phase velocity and wave number, respectively. In (c) and (d) the solid and dashed lines refer to the cases of real and imaginary values, respectively, for the phase velocity and wave number. Normalized parameters $\omega_1/\omega_0=2$ and $\omega_2/\omega_0=1$ are chosen for these plots.} \label{fig:set4}
\end{figure*}

\subsection{Membrane-based metamaterial including masses and springs}

As a final example of the hidden-force approach, consider a more general case of the previously analyzed membrane system obtained by including a mass and a spring in each unit cell attached to the membrane, as shown in Fig. 4(a). This  provides more degrees of freedom than even the previous example of the mass-and-spring analogy of the solid-matrix acoustic metamaterial. The applied
force $F=\Delta p S$ on a unit cell from the pressure gradient acts on the part of the unit-cell mass 
$M=\rho_0SD+M_{mem}$ made up as before of the sum of the masses of the air and the membrane. The hidden force $F_h=-k(\xi-\eta)-k_m\xi$ is the combination of the forces from the spring $-k(\xi-\eta)$, where $\xi$ and $\eta$
are the displacements of the masses $M$ and $M_c$, respectively, and the force $-k_m\xi$ from the membrane spring constant.
 The equations of motion for sinusoidal excitation of a single unit cell are $M\ddot{\xi}+k(\xi-\eta)+k_m\xi=
F_0\exp(-i\omega t)$ and $M_c\ddot{\eta}+k(\eta-\xi)=0$. The equations of motion are slightly different from the previous example of the mass-and-spring analogy of the solid-state matrix, but reduce to the same form when $k_m=0$. By elimination of the variable $\eta$ one can derive $M_{eff}$ in the following form:
\begin{equation} \label{seventz}
M_{eff}=F/\ddot{\xi}=M\left(1+\frac{\omega_1^2-\omega_0^2}{\omega_0^2-\omega^2}-\frac{\omega_2^2}{\omega^2}\right),
\end{equation}
where $\omega_0=\sqrt{k/M_c}$, $\omega_1=\omega_0\sqrt{1+M_c/M}$ and $\omega_2=\sqrt{k_m/M}$. For the special case in which the membrane spring constant $k_m$ can be neglected, we may set $\omega_2=0$. This results in the simpler form
\begin{equation} \label{sevent}
M_{eff}=M\left(1+\frac{\omega_1^2-\omega_0^2}{\omega_0^2-\omega^2}\right),
\end{equation}
which is precisely the same as Eq. (\ref{thirt}). In this special case, the present system is an exact analog of the mass-and-spring model previously considered. The effective modulus is still given by $B_0$ according Eq.~(\ref{three1}), because the pressure-volume relation, $p_j=-B_0\Delta V_j/V$, is not affected by the addition of the mass and spring.

Since $\rho_{eff}=M_{eff}/SD$ and $q=\omega\sqrt{\rho_{eff}/B_0}$, the dispersion relation is given by
\begin{equation} \label{eightt}
q=\omega\sqrt{\frac{\rho'}{B_0}}\left(1+\frac{\omega_1^2-\omega_0^2}{\omega_0^2-\omega^2}-\frac{\omega_2^2}{\omega^2},\right)^{1/2},
\end{equation}
where $\rho'=M/SD$. The  frequency spectra for the effective density and phase velocity, together with the dispersion relation, are shown in Fig. 4(b)-(d) for the case $\omega_1/\omega_0=2$ and $\omega_2/\omega_0=1$. For this choice of parameters, two distinct frequency bands with negative density are evident.

This mechanical model also provides the expected results in the limit when either the internal-spring constants $k/2$ go to infinity or the internal mass $M_c$ is set to zero. In both cases the model reduces to an elementary mass-and-spring chain model, which, in the metamaterial limit $D\ll\lambda$, shows a constant sound velocity (i.e no dispersion) and positive and constant effective mass.

To conclude this discussion of effective densities, we have proposed the hidden-force picture to explain why effective masses and
densities are significantly different from their non-resonant average values. In this picture the effective
mass $M_{eff}$ is obtained in terms of the hidden-force to applied-force ratio $F_h/F$ as
$M_{eff}=M/(1+F_h/F)$. The effective mass becomes negative when $F_h/F<-1$. We demonstrated that this
picture allows one to obtain effective masses of the unit cell and thereby the effective densities in a quick and easy manner for two established
metamaterials that exhibit negative density as well as for
a new membrane-based metamaterial including masses and springs.

In the next section we shall discuss with the aid of several examples a simple picture of how frequency-dependent effective moduli arise in acoustic metamaterials.

\section{Effective moduli: hidden sources or hidden expanders}

Here we introduce the concept of ``hidden sources'' of volume in order to understand effective modulus. This approach is first explained by considering the vibrational response of a piston connected to a chamber containing either a Helmholtz resonator or a side hole. We then consider an acoustic metamaterial based on unidirectional propagation in a tube lined with Helmholtz resonators. We go on to treat the case of a tube containing a combination of Helmholtz resonators and membranes\textemdash a generic case of a double-negative acoustic metamaterial\textemdash followed by a similar mass-and-spring analogy. For mass-and-spring models we extend the concept of hidden sources, more appropriately termed ``hidden expanders'' of displacement in this case, by the introduction of light rigid trusses coupled to extra degrees of freedom for mechanical motion, and demonstrate an example of a double-negative system based on this concept.  We conclude by summarizing our approach and discussing how to tell at first glance what produces effective density and what produces effective modulus. 

\subsection{Concept of a hidden source}

Consider a chamber and piston containing air as well as a point where air can be introduced or removed. This point, not apparent to the operator moving the piston, constitutes the origin of what we call a hidden source or sink of volume. A schematic diagram of this setup is shown in Fig. 5(a), where we represent a small change in which the piston is displaced to perturb the chamber volume by $\Delta V$ at the same time as a volume $\Delta V_h$ of air is introduced (measured at the equilibrium pressure before the change). The change in pressure $p$ inside the chamber is given by
\begin{equation} \label{ninet0}
p=-B_0\frac{\Delta V+\Delta V_h}{V}.
\end{equation}
The effective bulk modulus $B_{eff}$ only depends on the observable volume change $\Delta V$, so, in accord with the definition of Eq.~(\ref{eight}), $p=-B_{eff}\Delta V/V$ defines $B_{eff}$ for this system:
\begin{equation} \label{ninet}
B_{eff}=B_0\left(1+\frac{\Delta V_h}{\Delta V}\right).
\end{equation}
\begin{figure}
\begin{center}
\includegraphics*[width=1.0\columnwidth]{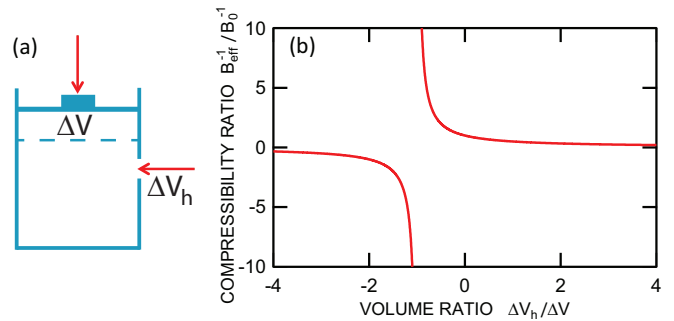}
\end{center}
\caption{(a) Showing the basic concept of a hidden source, in which volume $\Delta V_h$ of air is introduced into a chamber while depressing a piston to change the chamber volume by $\Delta V$. Here the changes as shown correspond to negative values of $\Delta V$ and $\Delta V_h$.  (b) shows a normalized plot of $B_{eff}^{-1}$ for this system as a function of the hidden source of  volume $\Delta V_h$.} \label{fig:set5}
\end{figure}
By analogy to the definition of $M_{eff}$ in Eq. (\ref{two}), $B_{eff}$ depends on the ratio of the hidden source $\Delta V_h$ to a more easily observable quantity, here the change in chamber volume $\Delta V$. A plot of $B_{eff}^{-1}/B_0^{-1}$ vs $\Delta V_h/\Delta V$ is shown in Fig. 5(b). Notably, as $\Delta V_h$ approaches $-\Delta V$, $B_{eff}$ becomes
zero (and compressibility $B_{eff}^{-1}$ becomes infinite). Also, $B_{eff}$ becomes negative when $\Delta V_h<-\Delta V$. This behavior is analogous to that observed in the case of hidden forces. (The curve in Fig. 5(b) is identical in shape to that in Fig. 1(c).)  Here one can see that a hidden source is represented by an introduced volume of air. 

\subsection{Single Helmholtz resonator}

A prototype of a system that can exhibit negative $B_{eff}$ is a piston and chamber with an attached Helmholtz resonator of volume $V_H$, as shown in Fig. 6(a). The neck of the resonator is assumed to have area $S_H$ and effective length\cite{blackstock2000fundamentals} $l'$. This system can be regarded as a piston with a chamber of equilibrium volume $V$, with pressure variations $p$ subject to a hidden
source of volume that is governed by the pressure variation $p_H=B_0S_H\eta /V_H$ inside the Helmholtz resonator, where $\eta$ is the displacement of the air plug of the resonator neck in the outward direction with respect to volume $V$. In the case in which the external driving force on the piston is sinusoidal at angular frequency $\omega$,   i.e. 
$p=p_0\exp (-i\omega t)$ and $\eta=\eta_0\exp (-i\omega t)$, the acceleration $\ddot{\eta}$ can be calculated from the equation of
motion, $\rho_0S_Hl'\ddot{\eta}=S_H(p-p_H)$, yielding
\begin{equation} \label{twent}
\eta_0=\frac{V_Hp_0}{B_0S_H}\frac{1}{1-\omega^2/\omega_0^2},
\end{equation}
where, for this case, $\omega_0=\sqrt{B_0S_H/(V_H\rho_0l')}$ is the classical Helmholtz resonator frequency.\cite{blackstock2000fundamentals} This treatment does not impose a limit on the ratio $V_H/V$, although, usually, $V_H<V$. The volume of the neck is, however, assumed to be much smaller than $V$ for the lumped-element treatment of the motion of the air inside it to apply. Higher-order resonances of the system are neglected in this approach.  
Knowing the displacement $\eta$ allows us to calculate the hidden source $\Delta V_h$:
\begin{equation} \label{twent1}
\Delta V_h=S_H\eta=\frac{V_Hp}{B_0}\frac{1}{1-\omega^2/\omega_0^2}.
\end{equation}
Using the harmonic expression $\Delta V=\Delta V_0\exp (-i\omega t)$, Eq. (\ref{twent1}), and the definition of $B_{eff}$ in Eq. (\ref{ninet}), we obtain, for the compressibility $B_{eff}^{-1}=-\Delta V_0/(Vp_0)$,
\begin{equation} \label{twent2}
B_{eff}^{-1}=B_0^{-1}\left(1+\frac{\omega_1^2-\omega_0^2}{\omega_0^2-\omega^2}\right),
\end{equation}
where $\omega_1=\omega_0\sqrt{(1+V_H/V)}$. The form of this equation is identical to that for $M_{eff}$ for the mass-and-spring analogy of the solid-matrix metamaterial and for the mass-and-spring membrane-based metamaterial in Eqs. (\ref{thirt}) and (\ref{sevent}), respectively. A plot of frequency vs $B_{eff}^{-1}$ is shown in Fig. 6(b) on a normalized scale for the case $\omega_1/\omega_0=2$. If the sinusoidal pressure amplitude $p_0$ is assumed be the imposed quantity, the volume amplitude $V_0$ is determined by $B_{eff}^{-1}$.
The volume will oscillate with large amplitude for $\omega$ approaching $\omega_0$ from below
because $B_{eff}^{-1}$ becomes very big.  In contrast, at $\omega=\omega_1$ when $B_{eff}^{-1}$=0 the system becomes infinitely rigid and the volume amplitude becomes zero. The abrupt shift of the phase of the volume variations by $\pi$ with respect to the driving pressure is also predicted as
the frequency passes through the resonance $\omega_0$. Since the sign of $B_{eff}$ changes at $\omega =\omega_0$, the region of negative $B_{eff}$ between $\omega_0$ and $\omega_1$ exhibits wave damping. Negative $B_{eff}$, a consequence of $\Delta V_h>-\Delta V$, thus implies in the case of sinusoidal excitation that the hidden source $\Delta V_h$ is not only in antiphase with but also has a magnitude larger than that of $\Delta V$. In the limit $\omega=0$, $B_{eff}=B_0(1+V_H/V)^{-1}$, which is reduced from the expected value $B_0$ owing to the increase in the total effective volume from $V$ to $V+V_H$. In the limit $\omega=\infty$, $B_{eff}=B_0$ because the flow of air to and fro from the Helmholtz resonator is effectively frozen owing to the inertia of the air plug in the resonator neck.

\begin{figure}
\begin{center}
\includegraphics*[width=1.0\columnwidth]{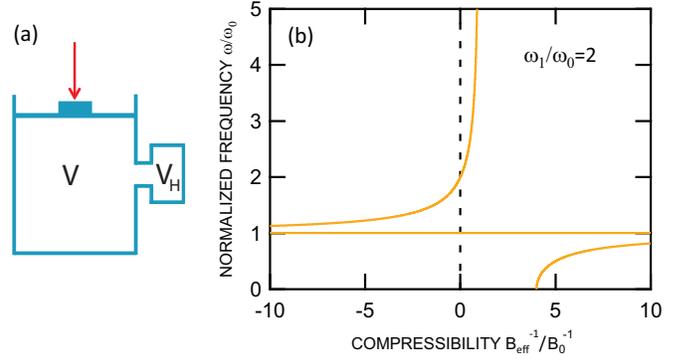}
\end{center}
\caption{(a) shows a prototype of a system that can exhibit negative $B_{eff}$, consisting of a piston and chamber with an attached Helmholtz resonator of volume $V_H$. (b) shows a normalized plot of frequency as a function of the compressibility $B_{eff}^{-1}$ for this system for the case $\omega_1/\omega_0=2$.} \label{fig:set6}
\end{figure}
\begin{figure}
\begin{center}
\includegraphics*[width=1.0\columnwidth]{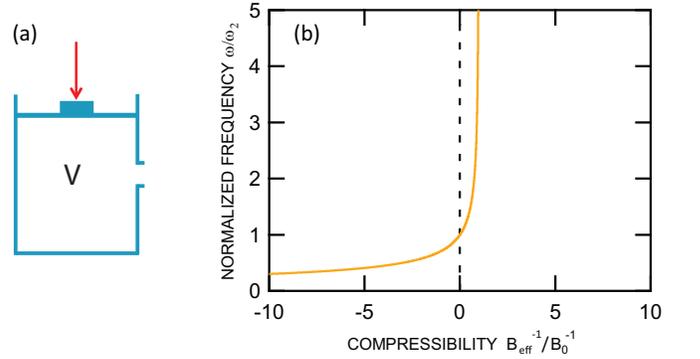}
\end{center}
\caption{(a) shows a prototype of a system that can exhibit negative $B_{eff}$, consisting of a piston and chamber with a side hole. (b) shows a normalized plot of frequency as a function of the compressibility $B_{eff}^{-1}$ for this system.} \label{fig:set7}
\end{figure}
The case of a side hole instead of a Helmholtz resonator, as shown in Fig. 7(a), is also one of practical interest in acoustic metamaterial design.\cite{lee2009acoustic,lee2010composite,lee2010reversed} For a side hole one may set $\omega_0=0$ because the Helmholtz resonator stiffness (i.e. spring constant), $k_H=S_HB_0/V_H$, vanishes. Equation (\ref{twent}) is modified to
\begin{equation} \label{twent3}
\eta_0=\frac{p_0}{\omega^2\rho_0l'},
\end{equation}
giving
\begin{equation} \label{twent4}
\Delta V_h=S_H\eta=-\frac{p}{\omega^2\rho_0l'}
\end{equation}
and
\begin{equation} \label{twent5}
B_{eff}^{-1}=B_0^{-1}\left(1-\frac{\omega_2^2}{\omega^2}\right),
\end{equation}
where $\omega_2=B_0S_H/(V\rho_0l')$. The resonance frequency $\omega_2$ depends on $V$ instead of $V_H$ in this case. Equation (\ref{twent5}) is the exact analog of Eq. (\ref{seven}) for the case of the effective mass of a membrane-based metamaterial. A normalized plot of the frequency dependence of $B_{eff}^{-1}$ is shown in Fig. 7(b). Negative $B_{eff}$ is exhibited up to $\omega=\omega_2$. In the limit $\omega=0$, $B_{eff}=0$, as expected  since the oscillating air is completely free to escape from the chamber in this case. In the limit $\omega=\infty$, $B_{eff}=B_0$ because the flow of air to and fro from the side hole is effectively frozen owing to the inertia of the air plug (as was the case with the Helmholtz resonator).

\subsection{Helmholtz-resonator-based acoustic metamaterial}

We are now in a position to consider the example of a 1D
acoustic metamaterial based on an array of Helmholtz-resonators spaced at regular intervals in a air-filled tube,\cite{fang2006ultrasonic,cheng2008one,hu2008homogenization,zhang2009focusing,lee2009acoustic,lee2010composite,lee2010reversed,seo2012acoustic,ding2010two,pope2010viscoelastic,fok2011negative,santillan2011acoustic,fey2011compact,seo2012acoustic,gracia2012double,chen2013double,yoo2014spatiotemporal,koju2014extraordinary,crow2015experimental} as shown schematically in Fig. 8(a). The unit cell consists of a section of cylindrical tube containing a single
Helmholtz resonator attached to the tube wall. 

\begin{figure*}
\begin{center}
\includegraphics*[width=2.0\columnwidth]{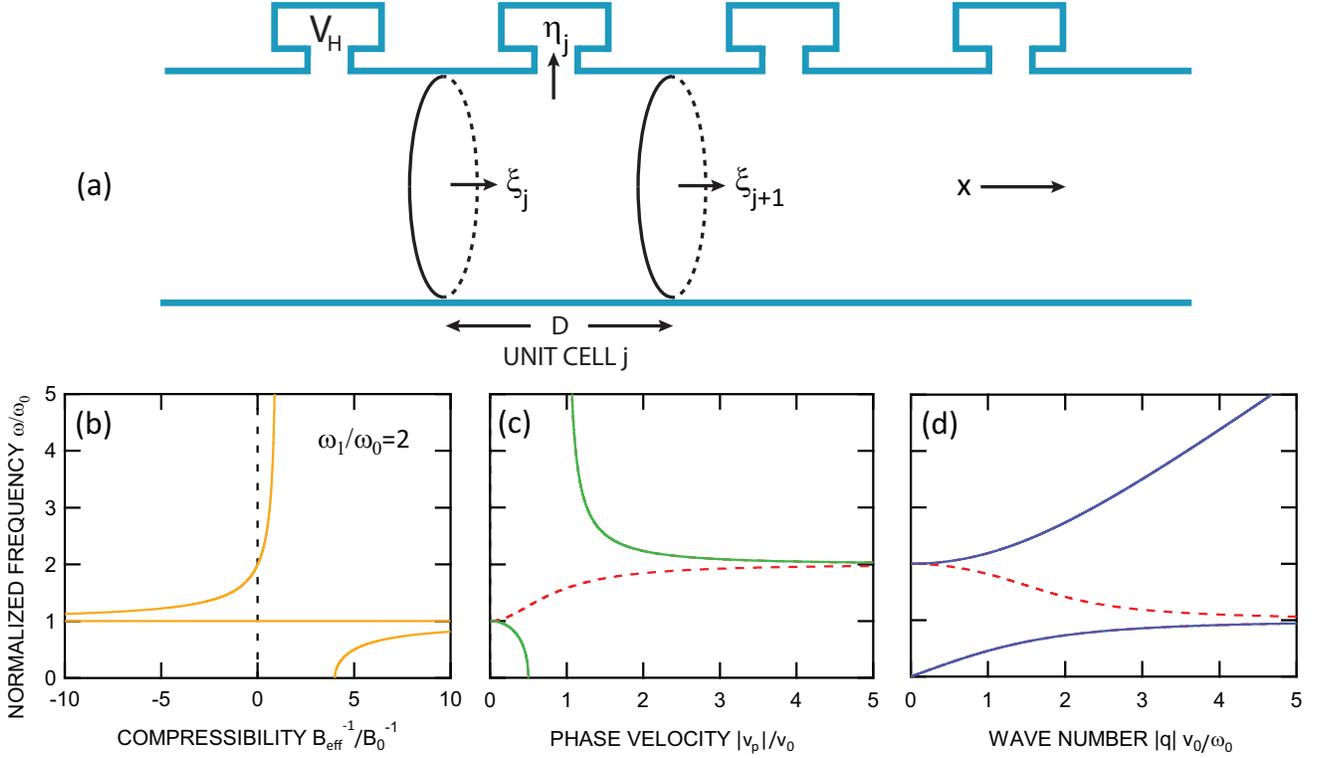}
\end{center}
\caption{(a) Schematic diagram of a 1D
Helmholtz-resonator-based acoustic metamaterial, showing periodically-spaced resonators in a tube. Normalized plots (b), (c), (d) for this system show the frequency as a function of the density, phase velocity and wave number, respectively, for the case $\omega_1/\omega_0=2$. In (c) and (d) the solid and dashed lines refer to the cases of real and imaginary values, respectively, for the phase velocity and wave number. In the horizontal axes we use velocity $v_0=\sqrt{B_0/\rho_0}$.} \label{fig:set8}
\end{figure*}

We first note that the effective density is equal to the density of air, $\rho_{eff}=\rho_0$, as is evident from the previously treated case of a tube containing membranes. (As there are no membranes here, the previously treated case, except with $k_m=0$, applies.) Here the unit cell volume change $\Delta V_j$ can be considered to depend on the non-equilibrium particle displacements $\xi_j$ and $\xi_{j+1}$ at the two unit cell boundaries, which act like pistons: $\Delta V_j=S(\xi_{j+1}-\xi_j)$. (The definition of $\xi_j$ here is distinct from that used for the membrane-metamaterial. Here it refers to the acoustic displacement at the left-hand boundary of the unit cell, rather than that of the center of mass of the cell.) However, in contrast to the situation for the membrane-based metamaterial [Eq. (\ref{eight2})], the average pressure change in the unit cell now also contains a contribution from the hidden source:
\begin{equation} \label{twent6}
p_j=-B_0\frac{\Delta V_j+\Delta V_{hj}}{V},
\end{equation}
which, in the present case, can be expressed as
\begin{equation} \label{twent6b}
p_j=-\frac{B_0}{SD}[S(\xi_{j+1}-\xi_j)+\eta_j S_H],
\end{equation}
where $\eta_j$ is the outward displacement of the Helmholtz-resonator air plug. From Eq. (\ref{ninet}), or, equivalently, using the definition
\begin{equation} \label{twent6a}
p_j=-B_{eff}\frac{S(\xi_{j+1}-\xi_j)}{SD},
\end{equation}
we obtain
\begin{equation} \label{twent7}
B_{eff}=B_0 \left(1+\frac{S_H\eta_j}{S(\xi_{j+1}-\xi_j)}\right).
\end{equation}
Introducing sinusoidally-varying quantities as before, making use of Eqs. (\ref{twent}), (\ref{twent6}) and (\ref{twent7}), and again assuming that $D\ll\lambda$, one again obtains Eq. (\ref{twent2}) for $B_{eff}^{-1}$. At the resonance frequency $\omega=\omega_1$, the effective modulus and the phase velocity become infinite. By analogy with zero-density metamaterials, this situation is useful for applications in extraordinary acoustic transmission.\cite{koju2014extraordinary,crow2015experimental}

To derive the dispersion relation in the limit $D\ll\lambda$, we make use of $\dot{p}=-B_{eff}\partial u/\partial x$ and $-\partial p/\partial x=\rho_{0}\dot{u}$ by analogy with Eqs.~(\ref{ten}) and (\ref{six}) to derive the wave equation:\footnote{As in all the derivations in this paper, we are making the assumption that vibrational amplitudes are small enough for any variations in $M_{eff}$ and $B_{eff}$ with time to be negligible.}
\begin{equation} \label{twent8}
\rho_{0}\ddot{p}=B_{eff}\frac{\partial^2 p}{\partial x^2},
\end{equation}
yielding $q^2=\omega^2\rho_0/B_{eff}$. For this example,
\begin{equation} \label{twent9z}
q=\omega \sqrt{\frac{\rho_0}{B_0}}\left(1+\frac{\omega_1^2-\omega_0^2}{\omega_0^2-\omega^2}\right)^{1/2}.
\end{equation}
The frequency dependences of the compressibility and the phase velocity, as well as the dispersion relation, are shown by normalized plots in Figs. 8(b)-(d). In the region of negative modulus the propagation is damped. This behavior is analogous to that noted for negative effective mass.

The equivalent result for the dispersion relation for an array of side holes instead of Helmholtz resonators is
\begin{equation} \label{twent9}
q=\omega \sqrt{\frac{\rho_0}{B_0}}\left(1-\frac{\omega_2^2}{\omega^2}\right)^{1/2},
\end{equation}
This has exactly the same form as the dispersion for the membrane-based metamaterial [Eq. (\ref{twelve})].

Just as one can ask the question why the addition of membranes produce no added rigidity for the passage of acoustic waves, one can also ask why Helmholtz resonators add no effective mass. In fact effective mass is only added, according to Eqs.~(\ref{two}) and (\ref{five1}), if there is an extra frequency-dependent force acting on a unit cell. Since there are no such hidden forces in the system of Helmholtz resonators but only hidden sources, these resonators do not contribute to the effective mass. Rather, the introduction of the Helmholtz resonators leads to an extra contribution $-B_0\Delta V_h/V$ to the pressure [Eq.~(\ref{twent6b})], whose effect on either side of a unit cell is the addition of a pair of equal and opposite forces. This pair of forces obviously does not contribute to an imbalance in the net force on a unit cell, as is required for the addition of effective mass [Eq.~(\ref{five1})], but instead leads to a net compression or expansion of the unit cell, i.e. resulting in a contribution to the elastic modulus.

\begin{figure*}
\begin{center}
\includegraphics*[width=1.5\columnwidth]{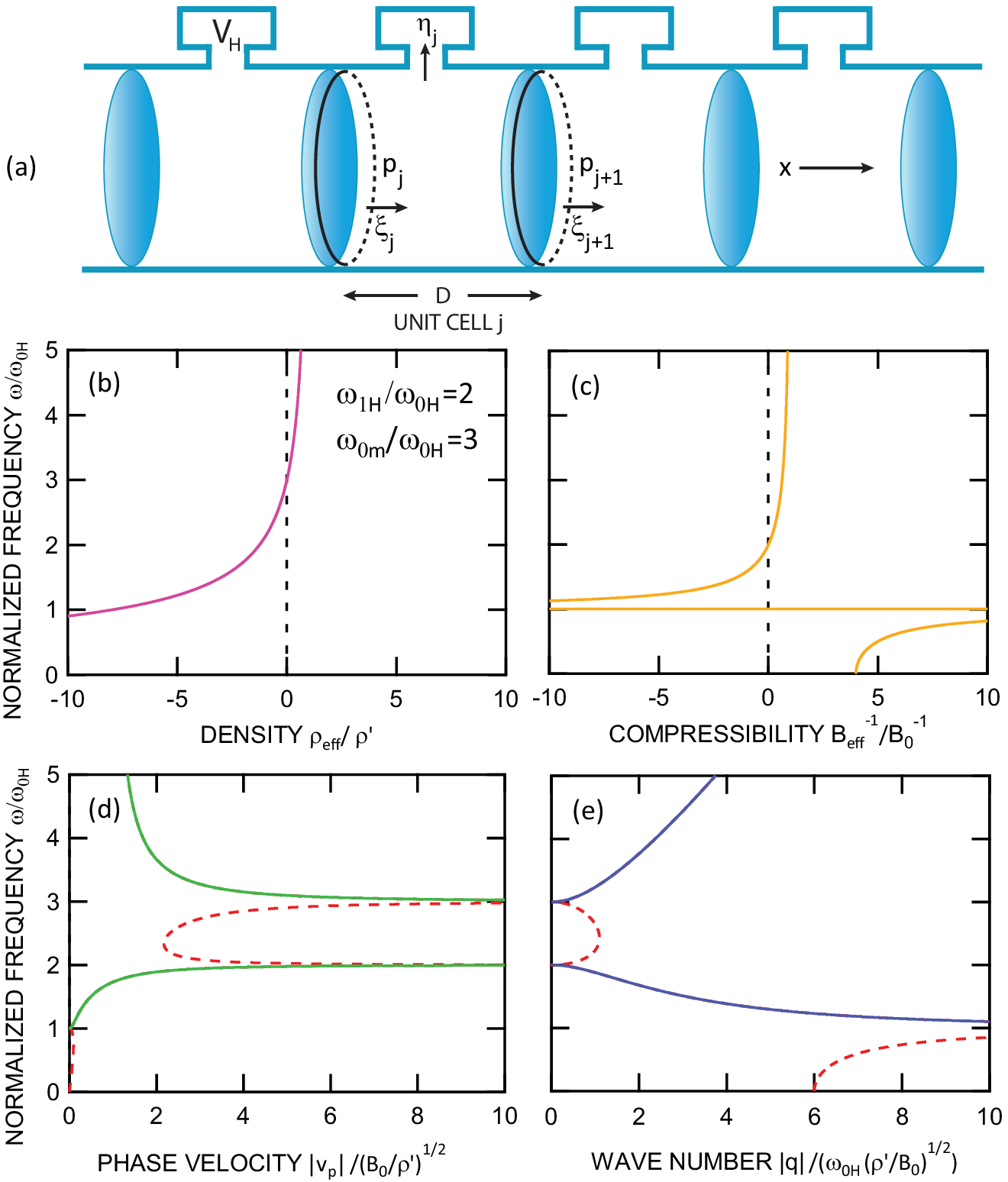}
\end{center}
\caption{(a) Schematic diagram of a 1D
acoustic metamaterial based on membranes and Helmholtz resonators, showing alternately-spaced elements in a tube. Normalized plots (b)-(e) for this system show the frequency as a function of the density, phase velocity and wave number, respectively, for the case $\omega_{1H}/\omega_{0H}=2$ and $\omega_{0m}/\omega_{0H}=3$, for which a region of double-negativity exists. In (d) and (e) the solid and dashed lines refer to the cases of real and imaginary values, respectively, for the phase velocity and wave number.} \label{fig:set9}
\end{figure*}

\subsection{Acoustic metamaterial with both Helmholtz resonators and membranes}

We now consider a 1D
acoustic metamaterial based on Helmholtz-resonators combined with membranes in an air-filled tube,\cite{lee2009acoustic,lee2010composite,lee2010reversed,seo2012acoustic} a prototype system for double-negative behavior, as shown in Fig. 9(a). The membranes and Helmholtz resonators are positioned alternately inside an air-filled tube. Consider the unit cell $j$ sketched in the dashed line in Fig. 9(a) that ends just after a membrane. This unit cell is chosen so it can apply to both the analysis of the membranes and Helmholtz resonators. The equation
\begin{equation} \label{four}
(p_j-p_{j+1})S-k_m\xi_j= M\dot{u_j},
\end{equation}
valid for the previously-treated case of membranes only, and
\begin{equation} \label{twent6b1}
p_j=-\frac{B_0}{SD}[S(\xi_{j+1}-\xi_j)+\eta_j S_H],
\end{equation}
i.e. Eq.~(\ref{twent6b}), derived for Helmholtz resonators only, still apply,
where the mass $M$ again refers to the lumped motion of the unit cell.
The change in the choice of unit cell affects the definition of the quantity $\xi_j$ in Eq.~(\ref{four}), which now refers to the acoustic displacement at the left-hand side of the unit cell $j$ rather than to that of the cell center of mass. However, the difference in these definitions, only affecting distances $\sim$$D/2$, does not lead to a change in the final results for effective physical properties and the dispersion relation. The above equations separately determine the effective density and modulus according to Eqs.~(\ref{five1}) and (\ref{three1}), so we arrive at expressions for $\rho_{eff}$ and $B_{eff}^{-1}$ in exactly the same form as those in Eqs. (\ref{seven}) and (\ref{twent2}), respectively:
\begin{equation} \label{thirt2}
\rho_{eff}=\rho'\left(1-\frac{\omega^2_{0m}}{\omega^2}\right),
\end{equation}
\begin{equation} \label{thirt3}
B_{eff}^{-1}=B_0^{-1}\left(1+\frac{\omega_{1H}^2-\omega_{0H}^2}{\omega_{0H}^2-\omega^2}\right),
\end{equation}
where we have added the labels $m$ for membrane and $H$ for Helmholtz resonator to remove the ambiguity in the definitions $\omega_{0m}=\sqrt{k_m/M}$, $\omega_{0H}=\sqrt{B_0S_H/(V_H\rho_0l')}$ and $\omega_{1H}=\omega_{0H}\sqrt{(1+V_H/V)}$.
By analogy with Eq. (\ref{fift}), the wave equation,
\begin{equation} \label{thirt4}
\rho_{eff}\ddot{\xi}=B_{eff}\frac{\partial^2 \xi}{\partial x^2},
\end{equation}
leads to the dispersion relation $q=\omega\sqrt{\rho_{eff}/E_{eff}}$, i.e.,
\begin{equation} \label{thirt5}
q=\omega \sqrt{\frac{\rho'}{B_0}}\left(1-\frac{\omega^2_{0m}}{\omega^2}\right)^{1/2}\left(1+\frac{\omega_{1H}^2-\omega_{0H}^2}{\omega_{0H}^2-\omega^2}\right)^{1/2}.
\end{equation}
The frequency dependence of the phase velocity, as well as the dispersion relation, are shown by normalized plots Fig. 9(d), (e) for the case $\omega_{1H}/\omega_{0H}=2$ and $\omega_{0m}/\omega_{0H}=3$, for which a region of double-negativity exists from $1<\omega/\omega_{0H}<2$. In this region the phase velocity is opposite to the group velocity. In 2D and 3D such materials are expected to be important in focusing applications. The dispersion relation for the case of side holes instead of Helmholtz resonators can be easily found by the use of Eq. (\ref{twent5}) for $B_{eff}^{-1}$ instead of Eq. (\ref{twent2}). We now turn to the case of a mass-and-spring model exhibiting an effective modulus or exhibiting both an effective density and modulus.

\subsection{Mass-and-spring analogy for an acoustic metamaterial exhibiting negative modulus or double-negative behavior}

Consider a 1D model consisting of masses and springs connected to light rigid hinged trusses coupled to extra degrees of freedom for mechanical motion, as shown in Fig. 10(a). We assume all mechanical displacements are much smaller than the truss lengths. This type of hinged truss system was previously proposed together with extra springs to generate a negative modulus,\cite{huang2012anomalous} but we have simplified the model to a convenient bare minimum here. Springs of constant $2k_0$ connect a mass $M$ in the unit cell to the truss systems. The square truss system, consisting of four members, is connected above and below to straight trusses, which in turn are connected to two masses $m$ that are constrained (by rails) to move only in the vertical direction. These ideal (i.e. massless and frictionless) hinged trusses ensure 1) that the same magnitude of force is exerted on the springs either side, and 2) that the vertical displacements $y$ of the masses $m$ are exactly mirrored by the horizontal displacements of the sides of the trusses attached to the springs. The presence of the two vertically-oriented straight trusses allows the square truss system to be free to move horizontally. The springs provide the applied force $F=F_j-F_{j+1}$ on a particular unit cell $j$, where, from Eq.~(\ref{five1}),
\begin{equation} \label{five11}
M_{eff}=\frac{F_j-F_{j+1}}{\ddot{\xi}_j}.
\end{equation}
However, from the force transmission properties of the truss system, that ensure that the compressive forces in the springs on either side of it are equal [see Fig. 10(a)], it is clear that the acceleration of mass $M$ is simply given by $\ddot{\xi}_j=(F_j-F_{j+1})/M$. Therefore, from Eq.~(\ref{five11}), $M_{eff}=M$ for this model.

\begin{figure*}
\begin{center}
\includegraphics*[width=1.5\columnwidth]{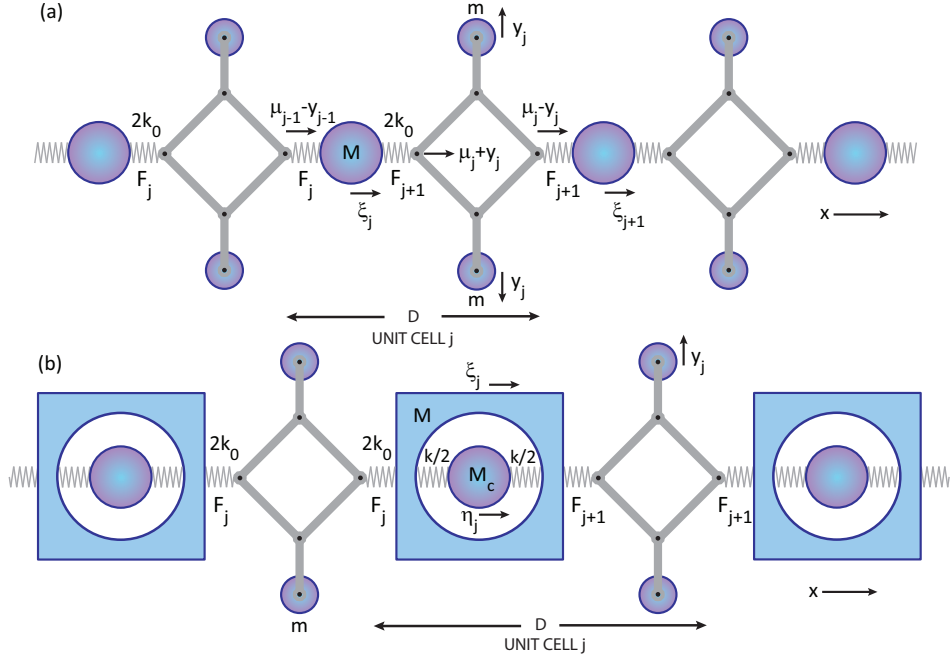}
\end{center}
\caption{(a) Mass-and-spring model that can exhibit negative modulus, realised by periodically-arranged masses and springs connected to light rigid trusses coupled to extra degrees of freedom for mechanical motion. (b) Mass-and-spring model that can exhibit double-negative behavior. $F_j$ and $F_{j+1}$ are the compressional forces in the springs on the left-hand side of and to the right of unit cell $j$, respectively. The ideal light rigid truss system containing two attached masses $m$ is hinged in such a way as to ensure that the compressional (or tensional) forces in the springs on either side of it are equal. The masses $m$ are constrained to move along fixed vertical rails.} \label{fig:set10}
\end{figure*}

To find the effective modulus, first consider the force $F_j$ on the left-hand side of unit cell $j$. This depends on the displacement $\mu_{j-1}-y_{j-1}$ of the right-hand side of the square truss of unit cell $j-1$ as well as on the displacement $\xi_j$ of mass $M$ in unit cell $j$, where $y_j$ and $-y_j$ are the displacements of the top and bottom masses $m$ in unit cell $j$, respectively, and $\mu_j$ is the horizontal displacement of the center of mass of the square portion of the truss system:
\begin{equation} \label{eight11}
F_j=2k_0(\mu_{j-1}-y_{j-1}-\xi_j).
\end{equation}
The compressional force $F_{j+1}$ in the spring to the right of mass $M$ in unit cell $j$ [see Fig. 10(a)] is given by
\begin{equation} \label{eight12}
F_{j+1}=2k_0(\xi_j-\mu_j-y_j).
\end{equation}
Decreasing the indices by 1 yields an alternative expression for $F_j$:
\begin{equation} \label{eight13}
F_{j}=2k_0(\xi_{j-1}-\mu_{j-1}-y_{j-1}).
\end{equation}
Comparing Eqs.~(\ref{eight11}) and (\ref{eight13}), we obtain
\begin{equation} \label{eight13a}
\mu_{j}=\frac{\xi_j+\xi_{j-1}}{2}.
\end{equation}
The massless square truss system moves by this amount to ensure equal displacements in the springs on either side of it and thus maintain the force balance on it. (Any massless system must by definition have a net zero force on it to avoid an infinite acceleration.) Eliminating $\mu_j$ from Eq.~(\ref{eight11})
\begin{equation} \label{eight14}
F_j=k_0(\xi_{j-1}-\xi_j-2y_{j-1}).
\end{equation}
From the properties of the truss system, force $F_j$ is transmitted to the masses $m$. For sinusoidal motion at angular frequency $\omega$, $F_{j}=-m\omega^2y_{j-1}$. This allows $y_{j-1}$ to be expressed as
\begin{equation} \label{eight15}
y_{j-1}=-\frac{\xi_{j}-\xi_{j-1}}{2(1-\omega^2/\omega_2^2)},
\end{equation}
where $\omega_2=\sqrt{2k_0/m}$ for this case. The definition of the effective Young's modulus, Eq.~(\ref{fourt1}), then leads to
\begin{equation} \label{fourt1a}
E_{eff}^{-1}=\frac{D}{k_0}\left(1-\frac{\omega_2^2}{\omega^2}\right).
\end{equation}
The effective modulus varies with frequency in exactly the same way as a tube containing an array of side holes [as in Eq.~(\ref{twent5})]. The frequency variation is the same as that shown in Fig. 7(b). So the model of Fig. 10(a) is the mechanical analog of an air-filled tube with periodically arranged side holes. At $\omega=0$ the effective modulus is zero because the truss system provides zero effective spring constant in this limit. In contrast to the mass-and-spring model of Fig. 3(b), the springs in the model of Fig. 10(a) cannot support any tension or compression in their equilibrium position, i.e. the springs should work in both tension and compression in the present case.\footnote{By the addition of a horizontal spring inside the truss system,\cite{huang2012anomalous} one can remove this constraint.} At $\omega=\infty$ the effective modulus becomes equal to $k_0/D$, identical to that of the previously considered mass-and-spring model, because in this limit the masses $m$ do not move and the square truss plays the role of a massless, rigid connector, resulting in two springs of constant $2k_0$ in series that are equivalent to a single spring of constant $k_0$. In this limit the effective modulus is simply $k_0/D$.

The question arises of how to interpret this mechanical model in terms of the hidden-source picture. The analogous equation to Eq.~(\ref{ninet0}) for the mass-and-spring model is, by comparison with the definition of Eq.~(\ref{eight1}),
\begin{equation} \label{fourt11}
F_j=-D^2\frac{k_0}{D}\frac{\xi_j-\xi_{j-1}}{D}-D^2\frac{k_0}{D}\frac{(\xi_j-\xi_{j-1})_h}{D},
\end{equation}
where $k_0/D$ is the modulus in the absence of the truss system and $(\xi_j-\xi_{j-1})_h$ is an extra displacement that we term a ``hidden expansion''. By comparing Eq.~(\ref{fourt11}) with Eqs.~(\ref{eight11}) and (\ref{eight15}), one finds that
$(\xi_j-\xi_{j-1})_h=2y_{j-1}$. In other words, the hidden expansion is precisely equal to the extra horizontal displacement introduced by the truss system. The mass-and-spring analogy of the hidden source concept is thus, quite naturally, a hidden expander.

The acoustic dispersion relation can be derived by analogy with the previously considered mass-and-spring model.
\begin{equation} \label{fourt12}
M\ddot{\xi_j}=2k_0(\mu_{j-1}-y_{j-1}-\xi_j)-2k_0(\xi_{j}-\mu_j-y_j),
\end{equation}
Provided that $D\ll\lambda$, this equation reduces to
\begin{equation} \label{fourt13}
M\ddot{\xi}=k_0D^2\frac{\partial^2 \xi}{\partial x^2}+2k_0D\frac{\partial y}{\partial x}.
\end{equation}
Compared to the previous case of Eq.~(\ref{fourt}), there is an extra term in $\partial y/\partial x$ owing to the truss system. A differential equation involving $\ddot{y}$ can be derived by noting that $m(\ddot{y}_j-\ddot{y}_{j-1})=F_{j+1}-F_j$, making use of Eqs.~(\ref{eight11}), (\ref{eight12}) and (\ref{eight13a}), assuming $D\ll\lambda$, and then integrating once over the coordinate $x$:
\begin{equation} \label{fourt14}
m\ddot{y}=-k_0D\frac{\partial \xi}{\partial x}-2k_0y.
\end{equation}
Substituting parameters with temporal variations according to $\exp[i(qx-\omega t)]$ as before leads to the dispersion relation $q=\omega\sqrt{\rho/E_{eff}}$, where $\rho=M/D^3$:
\begin{equation} \label{sixta}
q=\omega\sqrt{\frac{\rho D}{k_0}}\left(1-\frac{\omega_2^2}{\omega^2}\right)^{1/2},
\end{equation}
which has exactly the same form as Eqs.~(\ref{twelve}) and (\ref{twent9}).

Let us now turn to a more general 1D mass-and-spring model that can exhibit double-negative behavior, as illustrated in Fig. 10(b). We have combined the model of Fig. 10(a) with that of Fig. 3(b). The analysis proceeds in exactly the same way as for these two cases: $M_{eff}$ and $E_{eff}$ are given by Eqs.~(\ref{thirt}) and (\ref{fourt1a}), respectively, and the dispersion relation, $q=\omega\sqrt{\rho_{eff}/E_{eff}}$, becomes
\begin{equation} \label{sixts}
q=\omega\frac{\omega_0}{\omega_1}\sqrt{{\frac{\rho'D}{k_0}}}\left(1+\frac{\omega_1^2-\omega_0^2}{\omega_0^2-\omega^2}\right)^{1/2}\left(1-\frac{\omega_2^2}{\omega^2}\right)^{1/2}.
\end{equation}
Somewhat coincidentally, this has precisely the same form as that of Eq.~(\ref{thirt5}) for the case of an air-filled tube containing a periodic array of membranes and Helmholtz resonators. The frequency dependence of the term arising from the hidden expanders is the same as that for the membranes (which give rise to hidden forces), whereas the frequency dependence of the term arising from the hidden forces has the same form as that for the Helmholtz resonators (which give rise to hidden sources). The frequency spectra of $\rho_{eff}$ and $E_{eff}$ are analogous to the plots of Fig. 9(b), (c), the only difference being that the roles of the effective mass and modulus are reversed. The plots for $|v_p|$ and $|q|$ are the same as for Fig. 9(d) and (e) for equivalent dimensionless parameters.

We conclude this section by emphasizing that effective mass and modulus are not just theoretical constructs but also experimentally measurable quantities. According to their definitions in Eqs.~(\ref{five1}), (\ref{three1}) and (\ref{fourt1}), it suffices in principle to put pressure, force or displacement sensors at the appropriate points inside the metamaterial, and then the effective parameters can be experimentally derived. In contrast, merely measuring the dispersion relation, that depends on the combination $\rho_{eff}/B_{eff}$ or $\rho_{eff}/E_{eff}$, will not in general be sufficient to distinguish effective mass from effective modulus.
\section{Conclusions}

In conclusion, we have proposed the concepts of hidden force and hidden source of volume to respectively account for the effective densities and moduli of acoustic metamaterials. The superficially strange concepts of negative density and modulus are naturally accounted for in this picture when the hidden force/source operates in antiphase to and is bigger in magnitude than the force/volume-change engendered in its absence. We illustrate our approach in 1D for well-known air-based metamaterials involving membranes, Helmholtz resonators or side holes with the inclusion of the new case of an array of masses attached to membranes by springs. We also introduce examples based on generic mass-and-spring models, in which case the concept of a hidden source of volume is replaced by the concept of a hidden expander of displacement.

Deciding at first glance what contributes to an effective density or to an effective modulus depends on the system. For air-based acoustic metamaterials, membranes only contribute to the effective density whether or not they have an attached mass and spring. The reason for this is that they only involve local, time-dependent hidden forces that provide a net  force on the unit cell. A similar result applies to the mass-and-spring model representing a solid-matrix acoustic metamaterial based on heavy spheres in a soft matrix. In air-based acoustic metamaterials the effective modulus arises because of time-dependent hidden sources of air volume associated with Helmholtz resonators or side holes, that only produce pairs of forces acting equally and oppositely on either side of the unit cell (i.e. resulting in a zero net force on the unit cell). A similar result applies to the mass-and-spring models containing light rigid hinged trusses attached to masses. Although not discussed here, the inclusion of extra degrees of freedom in mass-and-spring or solid-state models, e.g. through rotations, can also result in an effective modulus through the production of pairs of equal and opposite forces.\cite{liu2011elastic,zhu2014negative, gusev2014double}

The extension of these ideas to 2D or 3D should be straightforward, whether for fluid, solid or multi-phase metamaterials.\cite{liu2011elastic} In these cases, due regard should be taken of possible anisotropic properties and of the different acoustic modes of propagation such as longitudinal, shear or flexural, for example. For simplicity in our treatment we have restricted our attention to  fundamental resonances, namely of membranes, Helmholtz resonators or side holes, although the inclusion of higher-order resonances in the general framework presented is possible. We have also ignored material damping. This has the advantage of highlighting the damping caused by the intrinsic metamaterial properties.  Frequency regions for single-negative-parameter behavior exhibiting damping can be referred to as band gaps, although they should not to be confused with those arising from purely phononic effects (generally observed at higher frequencies). Frequency regions for double-negative behavior are of particular interest because of their potential for high-resolution focusing. Finally, acoustic metamaterials will probably be entering the application stage as commercial products in sound control in the near future, and we hope that this paper will aid in accelerating progress in this regard.

\section*{Acknowledgements}

We are grateful to Alex Maznev, Vitalyi Gusev, Osamu Matsuda, Eun Bok, Insang Yoo, Jong Jin Park and Tomohiro Kaji for stimulating discussions. This work was supported by the Center for Advanced Meta-Materials (CAMM) funded by
the Ministry of Science, ICT and Future Planning as a Global Frontier Project, and by the
Basic Science Research Program through the National Research Foundation of Korea (NRF)
funded by the Ministry of Education, Science and Technology (CAMM-2014M3A6B3063712
and NRF-2013K2A2A4003469). We also acknowledge Grants-in-Aid for Scientific Research from the Ministry of Education, Culture, Sports, Science and Technology (MEXT) and well as support from the Japanese Society for the Promotion of Science (JSPS) 



%

\end{document}